\documentclass[aps,reprint,pra,letterpaper,floatfix,nobalancelastpage,notitlepage]{revtex4-1}
\usepackage{amsmath}
\usepackage{amsfonts}
\usepackage{graphicx}
\usepackage{bm}
\usepackage{color}
\usepackage{braket}
\usepackage{bbold}
\usepackage[utf8]{inputenc}
\usepackage{hyperref}

\begin{document}

\title{Efficient bit encoding of neural networks for Fock states}

\author{Oliver Kaestle}
\email{o.kaestle@tu-berlin.de}
\author{Alexander Carmele}
\affiliation{Technische Universit\"at Berlin, Institut f\"ur Theoretische Physik, Nichtlineare Optik und Quantenelektronik, Hardenbergstra{\ss}e 36, 10623 Berlin, Germany}
\date{\today}

\begin{abstract}
We present a bit encoding scheme for a highly efficient and scalable representation of bosonic Fock number states in the \textit{restricted Boltzmann machine} neural network architecture. In contrast to common density matrix implementations, the complexity of the neural network scales only with the number of bit-encoded neurons rather than the maximum boson number. Crucially, in the high occupation regime its information compression efficiency is shown to surpass even maximally optimized density matrix implementations, where a projector method is used to access the sparsest Hilbert space representation available.
\end{abstract}

\maketitle

\section{Introduction}

In recent breakthroughs, artificial neural networks have been successfully utilized for the description of quantum states~\cite{Carleo2017, DengSarma2017, DengSarmaX2017, Glasser2018, Torlai2018a, Schmitt2020, Burau2020, Nomura2017} and open quantum systems with Markovian dynamics~\cite{Yoshioka2019, Vicentini2019, Hartmann2019, Nagy2019}. In particular, the \textit{restricted Boltzmann machine} (RBM) neural network architecture has been established as a natural and highly efficient representation of the density matrix for spin and small molecular quantum systems~\cite{Carleo2017, Torlai2018, Melko2019, Amin2018, Hsieh2021, Xia2018, Alcalde2020, Sehayek2019, Huang2017}, as it allows for a one-to-one mapping of spins to artificial neurons and enables direct access to the stationary state via iterative application of a variational principle~\cite{Cui2015, Weimer2015}. While the implementation of periodic spin systems and spin systems with symmetries of translational invariance results in high numerical performance and fast convergence times~\cite{Choo2018, Yoshioka2019, Vicentini2019, Hartmann2019, Nagy2019, Yevick2021, Xiao2020}, adaptive strategies for the sampling of input system configurations have been shown to render accurate calculations of asymmetric open spin systems feasible as well~\cite{Kaestle2020sampling}.

In this article, we expand the representational power of the RBM architecture towards hybrid spin systems comprising bosonic Fock number states. To this end, a bit encoding scheme is applied to the Fock state basis, enabling a direct mapping of bosonic number states to the visible neuron layer without modification of the underlying neural network structure itself.
Strikingly, we find that in the regime of high Fock state occupation numbers, the bit-encoded neural network information compression efficiency surpasses even a maximally optimized density matrix representation in stationary state, where a projector method is employed to access the sparsest Hilbert space representation available. We demonstrate the accuracy of the presented neural encoding of Fock states by calculating the stationary boson number statistics of a generic one-atom laser model~\cite{Shore1993, Puri1986, Richter2009, Kreinberg2018, Gegg2018} and providing comparison benchmark calculations. Moreover, we demonstrate the methods' scalability potential into the large boson number regime where the information compression of the neural network becomes most efficient.
Aside from the goal of advancing the paradigm of the RBM as a universally applicable neural network architecture for the simulation of open quantum systems, specific applications of the presented method include, e.g., neural network realizations of boson sampling algorithms~\cite{Neville2017, Agresti2019} or of recent attempts to quantify quantum coherence via Fock state superposition~\cite{Lueders2021}.

The article is organized as follows: The investigated model system is introduced in Sec.~\ref{sec:model}. In Sec.~\ref{sec:bitencoding}, we derive the neural bit encoding scheme for Fock number states to enable a direct mapping to the visible neuron layer of the RBM. Afterwards, details on the implementation and the training procedure of the neural network are provided in Sec.~\ref{sec:training}.
In Sec.~\ref{sec:efficiency_gain}, the information compression efficiency of the bit-encoded RBM is compared to both a regular and a highly optimized density matrix implementation with respect to the required Fock state basis dimension. To this end, we compare the scaling of complexity for the considered model system, featuring a highly sparse Hilbert space in stationary state which can be truncated by making use of a Heisenberg projector method for maximum efficiency. Yet, in the regime of high Fock state occupations we find that the bit-encoded neural network still outperforms the competing approach with respect to compression efficiency.
Finally, in Sec.~\ref{sec:statistics} we demonstrate the accuracy of the presented method by calculating the stationary boson number statistics, before a confirmation of the methods' scalability potential for an accurate depiction of large Fock state occupations is provided in Sec.~\ref{sec:scalability}. Lastly, we summarize our findings in Sec.~\ref{sec:conclusion}.

%%%%%%%%%%%%%%%%%%%%%%

\section{Model} \label{sec:model}

To quantify the achievable information compression in systems comprising bosonic degrees of freedom via the presented bit-encoded neural network approach, we consider the paradigmatic open Jaynes-Cummings model, describing a realization of a one-atom laser via the interaction of a single spin system with a bosonic cavity mode~\cite{Kreinberg2018}.
In rotating wave and dipole approximation, the corresponding system Hamiltonian is given by~\cite{Shore1993, Puri1986, Richter2009, Gegg2018}
\begin{equation}
H / \hbar = \omega_0 \sigma^+ \sigma^- + \omega_c c^\dagger c + g_0 \left( \sigma^+ c + \sigma^- c^\dagger \right),
\label{eq:hamilton}
\end{equation}
with Pauli spin operators $\sigma^{\pm}$ and bosonic creation and annihilation operators $c^\dagger$, $c$. Here, $\omega_0$ and $\omega_c$ correspond to the spin and cavity mode frequencies and $g_0$ denotes the coupling amplitude between the system and cavity mode. In addition, the spin-$1/2$ system is incoherently driven at rate $\Gamma$, combined with an incoherent decay of the bosonic mode occupation at rate $\kappa$. The resulting time evolution dynamics for the density operator is prescribed by
\begin{equation}
\dot{\bm{\rho}} = \mathcal{\bm{L}} \bm{\rho} = -i \left[H/\hbar, \bm{\rho} \right] + \mathcal{\bm{D}} [\sqrt{\kappa/2} c] \bm{\rho} + \mathcal{\bm{D}} [\sqrt{\Gamma/2} \sigma^+] \bm{\rho}
\label{eq:JC_model}
\end{equation}
where we have introduced the Lindblad dissipators~\cite{Breuer2002, Mukamel1999}
\begin{subequations}
\begin{align}
\mathcal{\bm{D}} [\sqrt{\kappa/2} c ] \bm{\rho} &= \dfrac{\kappa}{2} \left( 2c \bm{\rho} c^\dagger - \{ c^\dagger c, \bm{\rho} \} \right), \\
\mathcal{\bm{D}} [ \sqrt{\Gamma/2} \sigma^+ ] \bm{\rho} &= \dfrac{\Gamma}{2} \left(2 \sigma^+ \bm{\rho} \sigma^- -  \{ \sigma^- \sigma^+, \bm{\rho} \} \right),
\end{align}
\end{subequations}
imposing incoherent excitation and dissipation on the system and the cavity mode, respectively. In the following calculations, we choose the parameters $g_0=0.2\,\mathrm{ps}^{-1}$, $\Gamma=0.4\,\mathrm{ps}^{-1}$, $\omega_0 = \omega_c$ and varying bosonic decay rates $\kappa$. Moreover, we are only interested in the stationary state reached at time $t_s$, where $\dot{\bm{\rho}}(t_s)=\mathcal{\bm{L}} \bm{\rho}(t_s)=0$ within numerical precision.

The corresponding system density matrix $\bm{\rho}$ consists of $2^d$ elements, with $d=2N n_\beta^{max}$ for a system comprising $N$ spins and a single bosonic mode with maximum occupation number $n_\beta^{max}$. In case of the here considered Jaynes-Cummings model, we have $N=1$. Due to the self-adjointness of the density matrix, only $d(d+1)/2$ of its elements must be determined for a complete system description.
Our model choice is motivated by the high sparsity of the stationary state density matrix: Using a Heisenberg projector method for maximum optimization, the full Hilbert space can be projected onto a subspace spanned by only $2(d-1)$ nonzero elements, completely describing the deterministic density matrix $\bm{\rho}(t_s)$ in stationary state~\cite{Fick1988, Breuer2002}.
In the following, we present a neural bit encoding scheme of Fock states based on the \textit{restricted Boltzmann machine} (RBM) neural network architecture. Here, the deterministic density matrix $\rho$ is estimated by a probabilistic neural density operator $\rho_{\bm{\vartheta}}$, which is fully described by a set of variational parameters $\bm{\vartheta}$. In the high boson number regime, the presented method is shown to yield a drastic reduction of complexity with respect to the deterministic density matrix representation, surpassing even the compression efficiency of the maximally optimized description.

\section{Neural encoding of Fock states} \label{sec:bitencoding}

The RBM neural network architecture can be employed to create a probabilistic model of the density matrix and is composed of binary neurons, meaning that each neuron in the network can take on one of two possible configurations. Recently, it has been shown to enable a highly favorable and efficient description of open spin systems via a one-to-one mapping of spins to binary neurons, establishing a natural representation of the systems' degrees of freedom~\cite{Carleo2017, Torlai2018, Melko2019, Choo2018, Hartmann2019, Nagy2019, Vicentini2019, Yoshioka2019, Carleo2019, Vieijra2020, Carleo2018, Cheng2018, Kaestle2020sampling, Rrapaj2021}.
The $2^{2N}$ density matrix elements $\bra{\sigma_1, \ldots, \sigma_N} \rho \ket{\eta_1,\ldots,\eta_N}$ for a system of $N$ spins $\sigma_n,\eta_n=\{-1,1\}$ are constituted by a model distribution referred to as \textit{neural density operator} which is optimized by iterative variation of a set of network parameters.
This neural network realization of the density matrix certainly is a great achievement, however, as of yet its potential has not been fully unleashed. In order to further expand the representational power of the RBM, in the following we present a highly efficient and scalable mapping of Fock number states to the artificial neurons by subjecting the bosonic Fock state basis to a bit encoding scheme~\cite{Kuhn2019, Kuhn2020}:

\begin{figure}[t]
\centering
\includegraphics[width=\linewidth]{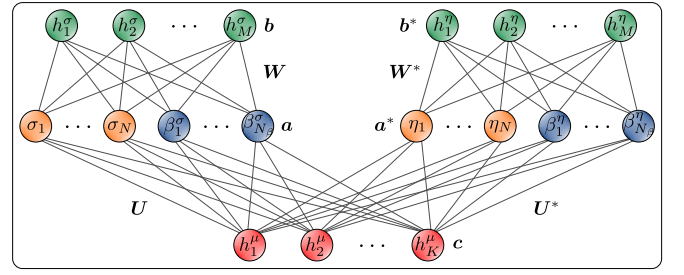}
\caption{RBM realization of the \textit{neural density operator}, featuring a visible layer storing the configuration of $N$ spin-$1/2$ systems (orange) and the bosonic Fock state occupation bit-encoded in $N_\beta$ neurons (blue), two hidden layers (green) and an ancillary mixing layer (red) with variational training parameters $\bm{\vartheta}=(\bm{a},\bm{b},\bm{c},\bm{W},\bm{U})$.}
\label{fig:rbm}
\end{figure}

The fundamental idea is to decompose the Fock state occupation number into a string of bits, which is then directly mapped onto the visible binary neurons of the RBM. To derive a general framework for hybrid systems comprising both spins and bosonic degrees of freedom, we consider $N$ spin-$1/2$ systems and a single bosonic mode, corresponding to density matrix elements $\bra{\sigma_1, \ldots, \sigma_N;n_\beta^\sigma} \rho \ket{\eta_1,\ldots,\eta_N; n_\beta^\eta}$ where $\sigma_n,\eta_n=\{-1,1\}$ again denote the left and right spin configurations and $n_\beta^\sigma,n_\beta^\eta \in \mathbb{N}_0$ correspond to the left and right number occupation of the bosonic mode.
The Fock state occupations $n_\beta^\sigma$, $n_\beta^\eta$ are each decomposed into $N_\beta$ bits $(\beta_1^\sigma,\ldots,\beta_{N_\beta}^\sigma)$ and $(\beta_1^\eta,\ldots,\beta_{N_\beta}^\eta)$, following the encoding rule
\begin{equation}
n_\beta = \sum_{i=1}^{N_\beta} 2^{i-1} \delta_{\beta_i,1},
\label{eq:bit_encoding}
\end{equation}
i.e., allowing for the representation of $n_\beta=\{0,1,\ldots,2^{N_\beta}-1 \}$ indistinguishable bosons on each side.
In this bit-encoded format, the Fock state basis can be directly mapped onto the binary neurons of the RBM analogous to the spin-$1/2$ systems and without any modification to the neural network architecture itself. Naturally, the regime of representable Fock state occupations is limited by the number of employed bits. For instance, utilizing a total of $N_{\beta}=4$ artificial neurons as bits corresponds to $2^4$ possible Fock state configurations in total, with the Fock occupation number given by
\begin{equation}
n_\beta = 2^0 \delta_{\beta_1,1} + 2^1 \delta_{\beta_2,1} + 2^2 \delta_{\beta_3,1} + 2^3 \delta_{\beta_4,1}.
\end{equation}

Fig.~\ref{fig:rbm} shows a sketch of the resulting bit-encoded RBM: The neural network features a visible layer of $2(N+N_\beta)$ sites $\bm{\sigma}=(\sigma_1, \ldots, \sigma_N, \beta_1^\sigma,\ldots,\beta_{N_\beta}^\sigma)$ and $\bm{\eta}=(\eta_1, \ldots, \eta_N, \beta_1^\eta,\ldots,\beta_{N_\beta}^\eta)$ representing the full configuration of the left and right side of the density matrix and consisting of $N$ spin-$1/2$ systems (orange shapes) and the bosonic mode occupation encoded in $N_\beta$ bits (blue shapes). In addition, the network comprises two auxiliary hidden layers with $M$ sites $\bm{h}^\sigma$ and $\bm{h}^\eta$ each (green shapes), connecting the visible sites of each side, and an ancillary mixing layer of $K$ neurons $\bm{h}^\mu$ connecting the left and right side of the density matrix (red shapes).
Tracing out the hidden and ancillary degrees of freedom, the elements of the neural density operator read~\cite{Hartmann2019, Nagy2019, Vicentini2019, Carleo2019, Kaestle2020sampling}
\begin{widetext}
\begin{align}
\rho_{\bm{\vartheta}} (\bm{\sigma}, \bm{\eta}) &= 8 \exp \Bigg[ \sum_{i=1}^{N} \left( a_i \sigma_i + a_i^* \eta_i \right) + \sum_{i=N+1}^{N+N_\beta} \left( a_{i} \beta_{i-N}^\sigma + a_{i}^* \beta_{i-N}^\eta \right) \Bigg]  \nonumber \\
& \times \prod_{m=1}^M \cosh \Bigg( b_m + \sum_{i=1}^N W_{mi} \sigma_i + \sum_{i=N+1}^{N+N_\beta} W_{mi} \beta_{i-N}^\sigma \Bigg)
\cosh \Bigg( b_m^* + \sum_{i=1}^N W_{mi}^* \eta_i + \sum_{i=N+1}^{N+N_\beta} W_{mi}^* \beta_{i-N}^\eta \Bigg) \nonumber \\
& \times \prod_{k=1}^K \cosh \Bigg[ c_k + c_k^* + \sum_{i=1}^N \left( U_{ki} \sigma_i + U_{ki}^* \eta_i \right) %
+ \sum_{i=N+1}^{N+N_\beta} \left( U_{ki} \beta_{i-N}^\sigma  + U_{ki}^* \beta_{i-N}^\eta \right) \Bigg],
\label{eq:rbm_ndo_traced}
\end{align}
\end{widetext}
where $\bm{\vartheta}=(\bm{a},\bm{b},\bm{c},\bm{W},\bm{U})$ denotes a set of complex training parameters split up into real and imaginary parts, yielding a total of $2(N+N_\beta) + 2M + K + 2M(N+N_\beta) + 2K(N+N_\beta)$ elements. These variational parameters constitute the networks' degrees of freedom, consisting of biases $\bm{a}$ for visible sites, $\bm{b}$ for hidden neurons and $\bm{c}$ for the mixing layer, and of complex weights $\bm{W}$ and $\bm{U}$ connecting the visible neurons $(\bm{\sigma},\bm{\eta})$ to the hidden layers $\bm{h}^\sigma$, $\bm{h}^\eta$ and to the ancillary mixing layer $\bm{h}^\mu$, respectively [see Fig.~\ref{fig:rbm}].

%%%%%%%%%%%%%%%%%%%%%%

\section{Training procedure} \label{sec:training}

Due to the exponential growth of the Hilbert space dimension with increasing system size, an exact mapping of the density matrix becomes increasingly expensive when considering large Fock state numbers. The artificial neural network ansatz approaches this problem by approximating the unknown density matrix $\rho$ by the neural density operator $\rho_{\bm{\vartheta}}$ [Eq.~\eqref{eq:rbm_ndo_traced}] via iterative optimization of the parameters $\bm{\vartheta}$. To this end, configuration space is efficiently compressed via application of the Metropolis algorithm~\cite{Metropolis1953}, where a sequence of $N_s$ samples of left and right density matrix configurations, i.e., visible neuron configurations of the RBM, is drawn as input data rather than taking every possible density matrix configuration into account. The Metropolis algorithm is based on a Markov chain Monte Carlo method, corresponding to a random walk in Hilbert space~\cite{Robert2004, Kampen2007, SchuldPetruccione2018}: A new system configuration $(\bm{\sigma}, \bm{\eta})=(\sigma_1, \ldots, \sigma_N, \beta_1^\sigma,\ldots,\beta_{N_\beta}^\sigma;\eta_1, \ldots, \eta_N, \beta_1^\eta,\ldots,\beta_{N_\beta}^\eta)$ is drawn based on the current sample and either accepted or rejected at a certain acceptance probability to find a subspace accurately representing the full Hilbert space of the considered system.
In many scenarios involving spin-$1/2$ systems interacting with bosonic modes, the number of nonzero combinations of spin configurations and Fock number occupations is severely limited by the structure of the spin-boson interaction, resulting in a highly sparse stationary state density matrix. Since our goal of training the neural network is to approximate only the steady state of the considered system, we exploit this fact to increase sampling efficiency and accuracy by only drawing samples from the subspace of nonzero steady state density matrix elements.
To propose a new sample, a random selection rule is employed where the left and right configuration of each spin $\sigma_1, \ldots, \sigma_N, \eta_1, \ldots, \eta_N$ is flipped at $50\%$ probability each. Afterwards, new random Fock number configurations $\beta_1^\sigma, \ldots, \beta_{N_\beta}^\sigma, \beta_1^\eta, \ldots, \beta_{N_\beta}^\eta$ are drawn based on the new spin configuration. Specifically, only combinations of spin configurations and boson numbers that have a nonzero stationary state contribution are taken into consideration as samples.
The acceptance probability of a newly drawn sample is chosen as
\begin{equation}
A(n+1,n) = \mathrm{min} \left[ 1, \dfrac{\tilde{p}_{\bm{\vartheta}}(\bm{\sigma}_{n+1},\bm{\eta}_{n+1})}{ \tilde{p}_{\bm{\vartheta}}(\bm{\sigma}_n,\bm{\eta}_n)} \right]
\end{equation}
where $(\bm{\sigma}_n,\bm{\eta}_n)$ denotes the current and $(\bm{\sigma}_{n+1},\bm{\eta}_{n+1})$ the newly proposed sample configuration.

Employing the stochastic reconfiguration approach~\cite{Sorella1997, Sorella2007, BeccaSorella2017}, the system observables and the normalized occurrence probability of a given sample configuration $(\bm{\sigma}_n,\bm{\eta}_n)$ with $n=\{1,\ldots,N_s\}$ are approximated as statistical expectation values over the $N_s$ samples drawn during one iteration. As a result, the \textit{normalized} occurrence probability is given by
\begin{equation}
\tilde{p}_{\bm{\vartheta}}(\bm{\sigma}_n,\bm{\eta}_n) = \dfrac{ | \rho_{\bm{\vartheta}} (\bm{\sigma}_n,\bm{\eta}_n) |^2 }{ \sum_{n=1}^{N_s} | \rho_{\bm{\vartheta}} (\bm{\sigma}_n,\bm{\eta}_n) |^2 },
\end{equation}
and diagonal observables can be estimated as statistical averages $\braket{X(\bm{\sigma},\bm{\sigma})} \approx \langle \langle X(\bm{\sigma},\bm{\sigma}) \rangle \rangle_q$~\cite{Sorella1997, Sorella2007, BeccaSorella2017, Hartmann2019, Vicentini2019, Nagy2019} with
\begin{equation}
\langle \langle X(\bm{\sigma},\bm{\sigma}) \rangle \rangle_q \! := \! \sum_{n=1}^{N_s} \tilde{q}_{\bm{\vartheta}}(\bm{\sigma}_n) \! \sum_{\bm{\xi}} \! X(\bm{\sigma}_n,\bm{\xi}) \dfrac{ \rho_{\bm{\vartheta}}(\bm{\xi}, \bm{\sigma}_n) }{ \rho_{\bm{\vartheta}} (\bm{\sigma}_n,\bm{\sigma}_n)},
\end{equation}
where we have introduced the normalized probability of diagonal system configurations $\tilde{q}_{\bm{\vartheta}}(\bm{\sigma}_n) = \rho_{\bm{\vartheta}} (\bm{\sigma}_n, \bm{\sigma}_n) / [\sum_{n=1}^{N_s} \rho_{\bm{\vartheta}} (\bm{\sigma}_n, \bm{\sigma}_n) ]$. In this work, we focus on diagonal observables as figures of merit. As a result, numerical performance can be further increased by employing the probability amplitude $\tilde{q}_{\bm{\vartheta}}(\bm{\sigma})$ based only on diagonal samples, which considerably reduces the dimension of the relevant configuration subspace: During each training iteration, $N_s$ diagonal samples $(\bm{\sigma}_n,\bm{\sigma}_n)$ are drawn to calculate $\tilde{q}_{\bm{\vartheta}}(\bm{\sigma})$ for the estimation of diagonal observables, and $N_s$ unrestricted samples $(\bm{\sigma}_n,\bm{\eta}_n)$ are drawn to calculate $\tilde{p}_{\bm{\vartheta}}(\bm{\sigma},\bm{\eta})$ for the training of the network.

The training goal is to determine the steady state of the considered system, prescribed by the condition $\dot{\rho} = \mathcal{L} \rho = 0$, with $\mathcal{L}$ denoting the Liouvillian superoperator~\cite{Breuer2002, Mukamel1999}. In order to optimize the parameter set $\bm{\vartheta}$ to fulfill this condition, we define a cost function $C(\bm{\vartheta}) = \left\| \mathcal{L} \rho_{\bm{\vartheta}} \right\|_2^2$~\cite{Vicentini2019, Nagy2019}. Initially, the variational parameters are set to small nonzero random values, $\vartheta_l^{(0)} \in [-0.01,0.01] \backslash \{ 0\}$. Using the standard stochastic gradient descent algorithm and $N_s$ sample system configurations as input training data, during each training iteration $t \rightarrow t+1$ the parameters $\bm{\vartheta}$ are updated by the rule
\begin{equation}
\vartheta_l^{(t+1)} = \vartheta_l^{(t)} - \nu \nabla_{\vartheta_l} C[\bm{\vartheta}^{(t)}],
\end{equation}
at a learning rate $\nu$~\cite{SchuldPetruccione2018}.
The required cost function gradient is evaluated as~\cite{Vicentini2019, Kaestle2020sampling}
\begin{align}
&\nabla_{\vartheta_l} C(\bm{\vartheta}) = 2 \mathrm{Re} \Bigg\{ \sum_{n=1}^{N_s} \tilde{p}_{\bm{\vartheta}} (\bm{\sigma}_n,\bm{\eta}_n) \tilde{\mathcal{\bm{L}}}^\dagger (\bm{\sigma}_n,\bm{\eta}_n) \nonumber \\
&\times \sum_{m=1}^{N_s} \mathcal{\bm{L}} (\bm{\sigma}_n,\bm{\eta}_n, \bm{\sigma}_m, \bm{\eta}_m) \dfrac{\bm{\rho}_{\bm{\vartheta}}(\bm{\sigma}_m, \bm{\eta}_m)}{\bm{\rho}_{\bm{\vartheta}}(\bm{\sigma}_n, \bm{\eta}_n)} O_{\vartheta_l}(\bm{\sigma}_m, \bm{\eta}_m) \nonumber \\
&- \left[ \sum_{n=1}^{N_s} \tilde{p}_{\bm{\vartheta}} (\bm{\sigma}_n,\bm{\eta}_n) O_{\vartheta_l}(\bm{\sigma}_n, \bm{\eta}_n) \right] \nonumber \\
& \times \left[ \sum_{n=1}^{N_s} \tilde{p}_{\bm{\vartheta}} (\bm{\sigma}_n,\bm{\eta}_n) \tilde{\mathcal{\bm{L}}}^\dagger (\bm{\sigma}_n,\bm{\eta}_n) \tilde{\mathcal{\bm{L}}} (\bm{\sigma}_n,\bm{\eta}_n) \right]  \Bigg\},
\end{align}
introducing the estimator of the Liouvillian 
\begin{equation}
\tilde{\mathcal{L}}( \bm{\sigma}_n, \bm{\eta}_n) \! := \! \! \sum_{\bm{\sigma}_m,\bm{\eta}_m} \mathcal{L} (\bm{\sigma}_n, \bm{\eta}_n, \bm{\sigma}_m, \bm{\eta}_m ) \dfrac{ \rho_{\bm{\vartheta}} (\bm{\sigma}_m,\bm{\eta}_m) }{ \rho_{\bm{\vartheta}} (\bm{\sigma}_n,\bm{\eta}_n)},
\end{equation}
and logarithmic derivatives stored in diagonal matrices with elements
\begin{equation}
[\bm{O}_{\vartheta_l}]_{\bm{\sigma}_n \bm{\eta}_n, \bm{\sigma}_n \bm{\eta}_n} = O_{\vartheta_l} (\bm{\sigma}_n, \bm{\eta}_n) = \dfrac{ \partial [ \ln \rho_{\bm{\vartheta}}(\bm{\sigma}_n,\bm{\eta}_n) ]}{ \partial \vartheta_l},
\end{equation}
which correspond to the neural density operator gradients with respect to all $l$ elements of $\bm{\vartheta}$ and for a given sample configuration $(\bm{\sigma}_n,\bm{\eta}_n)$.

%%%%%%%%%%%%%%%%%%%%%%

\section{Neural network efficiency gain} \label{sec:efficiency_gain}

In a regular density matrix implementation, the number of required elements for a complete system description scales polynomially with the maximum boson number $n_\beta^{max}$. For the here considered model [Eq.~\eqref{eq:JC_model}], this corresponds to $2n_\beta^{max} (2n_\beta^{max} + 1)/2$ elements, with $n_\beta^{max}$ denoting the chosen bosonic occupation number limit dictated by the numerical implementation. In its maximally optimized stationary state representation, a linear scaling via $2(2n_\beta^{max}-1)$ can be achieved.
In contrast, in the presented bit-encoded neural network the amount of variational parameters arising from bosonic degrees of freedom scales only with the number of \textit{bits} $N_\beta$, with $n_\beta^{max}=2^{N_\beta}-1$, corresponding to a drastic decrease of complexity especially in the limit of large boson numbers.

%%%%%%%%%%%%%%%%%%%

\begin{figure}[t]
\centering
\includegraphics[width=\linewidth]{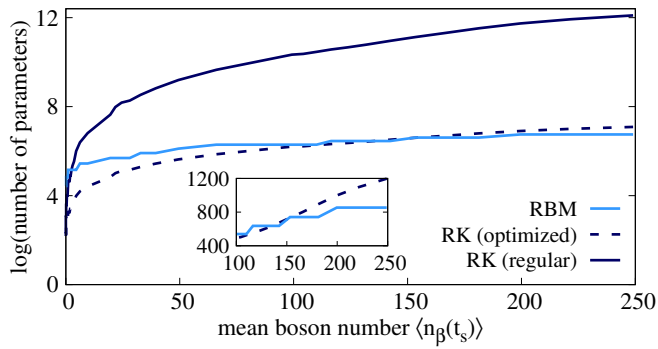}
\caption{Required number of parameters for a complete and numerically convergent system description, plotted on a logarithmic scale with respect to the mean stationary state boson occupation. The RBM approach (solid light blue line) is compared to a regular density matrix implementation (solid dark blue line) and a highly optimized approach with a truncated Hilbert space featuring only nonzero steady state density matrix elements (dashed dark blue line). Inset: Cutout on a linear scale, showing the area where the RBM implementation becomes the most efficient.}
\label{fig:scaling}
\end{figure}

In Fig.~\ref{fig:scaling}, we compare the number of parameters required for a complete and numerically convergent description of the considered model system with respect to the average boson occupation number in stationary state $\braket{n_\beta(t_s)}$ and plotted on a logarithmic scale. A lower value corresponds to a higher degree of information compression. The mean stationary Fock state occupation is tuned by variation of the bosonic decay rate $\kappa$.
In the neural network implementation, convergence is achieved once the number of employed bits $N_\beta$ is chosen sufficiently large and can be further improved by increasing the number of samples per iteration $N_s$. Numerical convergence of the regular density matrix implementation is assumed if further expanding the maximum Fock state occupation $n_\beta^{max}$ results in a relative deviation of less than $0.1\%$ in $\braket{n_\beta(t_s)}$.
With increasing degrees of freedom, dynamical Runge Kutta calculations typically require an increasingly small time discretization to achieve numerical convergence. In addition, the required number of elements scales \textit{polynomially}, resulting in a polynomial increase in complexity for rising system sizes (solid dark blue line).
Exploiting the sparsity of the stationary state density matrix to truncate the corresponding Hilbert space via application of a projector method, the density matrix implementation can be maximally optimized to scale \textit{linearly} in the required number of parameters (dashed dark blue line).
The number of variational RBM parameters defining the neural density operator scale with the number of employed bits $N_\beta$. While increasing the hidden layer sizes of course results in a less efficient compression, we note that in our experience numerical convergence of the network can be improved a lot more efficiently by increasing the bosonic degrees of freedom $N_\beta$ rather than the hidden layer dimensions $M$ and $K$. Therefore, the solid light blue line in Fig.~\ref{fig:scaling} shows the required number of variational parameters to achieve a convergent estimation of the density matrix at fixed hidden layer densities $M/(N_\beta+1)=K/(N_\beta+1)=1$, exhibiting a slow linear increase for rising Fock state basis dimensions.

As a main result of our study, Fig.~\ref{fig:scaling} illustrates a much more efficient compression of system information by the RBM architecture with respect to the regular density matrix implementation. The inset shows a cutout on a linear scale, where the bit encoding of the bosonic degrees of freedom results in a stepwise increase of complexity (solid light blue line).
The maximally optimized, linearly scaling density matrix implementation is comparably efficient and even undercuts the required number of variational RBM parameters in the low Fock state occupation regime. Strikingly, the neural network information compression becomes even more efficient above $\braket{n_\beta(t_s)} \approx 160$ (see inset). Given the already excellent Hilbert space compression achieved by the projector method in the maximally optimized density matrix approach, this is a remarkable result.
In the following, we explicitly demonstrate the bit-encoded RBMs' accuracy and scalability potential with regard to the regime of large Fock state basis dimensions.

\begin{figure}[t]
\centering
\includegraphics[width=\linewidth]{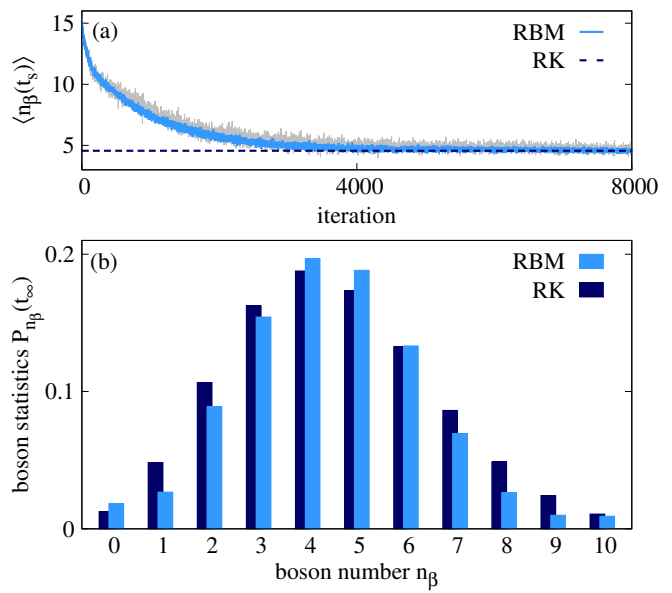}
\caption{Demonstration of the accuracy of the bit-encoded neural network implementation of Fock number states. (a) Expectation value for the stationary Fock state occupation number $\braket{n_\beta(t_s)}$ obtained from the RBM (solid blue line), compared to a calculation using fewer samples per iteration (solid grey line) and to the benchmark result (dashed line). (b) Steady state boson number statistics resulting from the RBM implementation (light blue bars) and compared to benchmark results (dark blue bars).}
\label{fig:photon}
\end{figure}

\section{Accuracy} \label{sec:statistics}

As a proof of principle and to demonstrate the accuracy of the neural encoding of Fock states, we specifically calculate the stationary boson occupation number statistics $P_n(t_s)$ for the considered model system [Eq.~\eqref{eq:JC_model}], with
\begin{equation}
P_n (t) = \dfrac{1}{n!} \braket{c^{\dagger n}c^n(t)} - \frac{1}{n!} \sum_{m=1}^{n^{max}} \dfrac{(n+m)!}{m!} P_{n+m}(t)
\label{eq:photon_statistics}
\end{equation}
denoting the probability of measuring $n$ bosons in the system at a given time $t$, calculated up to the highest included bosonic correlation degree $n^{max}$~\cite{Kabuss2012, Kabuss2011a}. Here we choose a low bosonic decay rate $\kappa=0.04\,\mathrm{ps}^{-1}$. In accordance with Fig.~\ref{fig:scaling}, we have chosen $N_\beta=5$ bits and hidden layer densities $M/(N_\beta+1)=K/(N_\beta+1)=1$ to achieve numerically convergent results. Calculations are performed at a learning rate $\nu=0.01$ and using $N_s=5000$ sample configurations per iteration. As a benchmark, we additionally calculate the system dynamics up to the steady state using a common density matrix implementation using identical parameters, $n_\beta^{max}=14$ and a time discretization $\Delta t = 0.02\, \mathrm{ps}$.

Fig.~\ref{fig:photon}(a) shows the estimated stationary state expectation value of the Fock state occupation number $\braket{n_\beta(t_s)}$ with respect to the number of training iterations of the RBM (solid light blue line) and compared to the benchmark result $\braket{n_\beta(t_s)} \approx 4.56$ (dashed dark blue line), exhibiting excellent agreement after approximately $4000$ iterations. The light oscillating behavior of the RBM results can be further reduced by increasing the number of samples per iteration $N_s$: Accordingly, a comparison RBM calculation using five times fewer samples per iteration exhibits increased variations (solid grey line).
Fig.~\ref{fig:photon}(b) depicts the steady state boson number statistics $P_{n_\beta(t_s)}$ [Eq.~\eqref{eq:photon_statistics}] calculated via training of the neural network (light blue bars) and compared to benchmark results (dark blue bars). The two resulting statistics are in overall very good qualitative agreement, sharing their highest boson number probability at $n_\beta=4$, with a Kullback-Leibler divergence of approximately $0.14$ which can be further reduced by increasing the sample size $N_s$.
It is noted, however, that the statistics resulting from the RBM implementation is prone to error accumulation for $n_\beta > 10$: The estimated occurrence probabilities feature statistical deviations arising from the Monte Carlo sampling procedure. These deviations are relatively small when considering the boson number observable $\braket{n_\beta}=\braket{c^\dagger c}$ and choosing a sufficiently large sample size $N_s$ [solid light blue line in Fig.~\ref{fig:photon}(a)]. However, during the calculation of Eq.~\eqref{eq:photon_statistics} the statistical error multiplies for each increasing correlation order $n$ of $\braket{c^{\dagger n} c^n}$,  thus limiting high accuracy RBM calculations of the boson number statistics to the low boson number regime for the considered sample size.

\section{Scalability} \label{sec:scalability}

\begin{figure}[t]
\centering
\includegraphics[width=\linewidth]{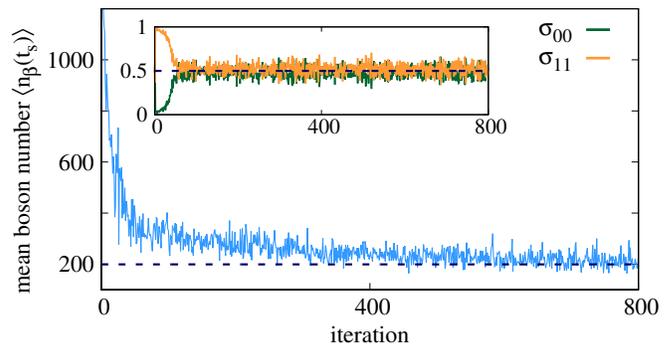}
\caption{Demonstration of the scalability potential of the bit-encoded neural network in the large Fock number regime, showing the mean Fock occupation number $\braket{n_\beta(t_s)}$ over training iterations (solid light blue line). The inset shows the spin down (green bottom line) and spin up occupations (orange upper line) of the spin system over iterations. Dashed dark blue lines indicate corresponding benchmark results.}
\label{fig:highscale}
\end{figure}

To access the high boson number regime, we calculate the considered model system [Eq.~\eqref{eq:JC_model}] once more at a small bosonic decay rate $\kappa=0.001\,\mathrm{ps}^{-1}$, resulting in $\braket{n_\beta(t_s)} \approx 199$ where the information compression efficiency of the RBM realization has been shown to surpass even a maximally optimized density matrix implementation (see Fig.~\ref{fig:scaling}). For training, we employ $N_\beta=13$ bits and hidden layer densities $M/(N_\beta+1)=K/(N_\beta+1)=1$ at a learning rate $\nu=0.003$ and $N_s=5000$ samples per iteration.
Even though $\braket{n_\beta(t_s)}$ is located well below the maximum Fock state occupation $n_\beta^{max}=2^{N_\beta}-1$, choosing fewer bits $N_\beta$ yields non-converging results, underlining the networks' need for sufficient degrees of freedom to facilitate effective training~\cite{Sehayek2019}. Thanks to the favorable scaling of the required number of variational parameters with increasing system sizes, calculations are still highly efficient in this regime.
Fig.~\ref{fig:highscale} shows the neural network results for the mean Fock state occupation number $\braket{n_\beta(t_s)}$ over the course of training iterations (solid blue line). Remarkably, already after approximately $400$ iterations, it approaches the benchmark value $\braket{n_\beta(t_s)} \approx 199$ (dashed blue line). The inset shows the steady state spin up and spin down expectation values of the single spin system obtained from the RBM implementation (blue and orange lines) and in good agreement with their corresponding benchmark results (dashed blue lines).
To conclude, the required number of neurons employed as bits $N_\beta$ to account for bosonic degrees of freedom exceed the actual stationary boson occupation by far. However, the number of training iterations to achieve numerical convergence is drastically reduced with increasing neuron numbers. This can be explained by the decreased asymmetry of the spin-boson interaction [Eq.~\eqref{eq:hamilton}] in the large boson number regime $n \gg 1$ where $\sqrt{n} \approx \sqrt{n+1}$, since the RBM architecture is known to achieve far higher levels of performance and convergence for the representation of systems with symmetries of translational invariance~\cite{Kaestle2020sampling}. At the same time, the bit-encoded neural network performs more efficiently than even highly optimized common implementations where a projector method has been employed to access the sparsest Hilbert subspace available, underlining the performance of the bit-encoded neural network representation of Fock states in the high occupation regime.

%%%%%%%%%%%%%%%%%%%%%%

\section{Conclusion} \label{sec:conclusion}

We have presented a bit-encoded realization of Fock number states in the RBM neural network architecture, extending its applicability of high-performing approximate mappings of the density matrix to hybrid spin systems featuring bosonic degrees of freedom, further advancing the paradigm of a universally applicable neural network architecture for open quantum systems.
Crucially, in the limit of large Fock state occupation numbers the RBM implementation requires severely fewer parameters for a complete system description than common density matrix approaches and even surpasses the information compression efficiency of a maximally optimized implementation, where the corresponding Hilbert space has been truncated to the sparsest possible representation by application of a projector method.
We have demonstrated the accuracy of the presented neural encoding of Fock states by calculating the stationary state boson number statistics of a model system, exhibiting good agreement with benchmark calculations. Moreover, to illustrate the scalability potential and performance of our method, we have calculated the mean stationary Fock state occupation in the high boson number regime, where the information compression of the neural network becomes the most efficient. Once numerical convergence is achieved by tuning the number of visible neurons in the network it can be further improved, e.g., by increasing the number of samples per iteration or via application of adaptive sampling strategies~\cite{Kaestle2020sampling}.

%%%%%%%%%%%%%%%%%%%%%%

\begin{acknowledgments}
We thank Marten Richter for fruitful discussions. The authors acknowledge support from the Deutsche Forschungsgemeinschaft (DFG) through SFB 910 project B1 (Project No. 163436311).
\end{acknowledgments}

%%%%%%%%%%%%%%%%%%%%%%

%

\end{document}